\documentclass[lettersize,journal]{IEEEtran}
\usepackage{amsmath,amsfonts}
\usepackage{algorithmic}
\usepackage{algorithm}
\usepackage{array}
\usepackage[caption=false,font=normalsize,labelfont=sf,textfont=sf]{subfig}
\usepackage{textcomp}
\usepackage{stfloats}
\usepackage{url}
\usepackage{verbatim}
\usepackage{graphicx}
\usepackage{cite}
\begin{document}

\title{Low Power 16-channel Wave Union TDC in a Radiation Tolerant FPGA}

\author{Brian A. Bryce, Kathryn M. Marcotte}%



\maketitle

\begin{abstract}
The design and performance of wave union TDC implemented in a Lattice CertusPro-NX FPGA is discussed. This FPGA is available for radiation tolerant applications. The TDC is implemented with 16-channels and a 200 MHz reference clock. Each channel is able to record at an event rate of \textgreater 1 MHz. The performance of the TDC is assessed over voltage and temperature. Typical TDC performance has a resolution of 10.9 ps. Typical INL is +/-3 LSB peak-to-peak. Typical DNL is (+1.13,-0.77) LSB. Typical differential performance between two channels is 20 ps (1-sigma).
\end{abstract}

\begin{IEEEkeywords}
time-to-digital-converter (TDC), time-of-flight (ToF), FPGA.
\end{IEEEkeywords}

\section{Introduction}
\IEEEPARstart{T}{ime-to-digital} converters (TDCs) are critical components for time-of-flight (ToF) based space instrumentation. Examples include energetic neutral atom cameras \cite{hena}, ion spectrometers \cite{psp, eps}, UV spectrometers \cite{UV}, and LiDARs \cite{altimeter}. Energetic particle instrumentation is the primary use case for the present work. These instruments make use of TDCs for both ToF and physical position interpolation via delay lines.

Many FPGA based TDCs have been developed \cite{TDCreview}. These designs are typically based on commercial grade FPGAs of a variety of types, and often focus on maximizing time precision. In this work we have selected a family of FPGA that is available as a radiation tolerant part built on 20 nm fully-depleted silicon-on-insulator (FD-SOI), the Lattice CertusPro-NX. This process makes this device both lower power and more radiation tolerant. For this work we made use of LFCPNX-100-9BBG484I, a screened version of this part is available in a lot controlled flow from Frontgrade as UT24CP1008, which is radiation tolerant to a TID of 100 krad. The modest fabric size of this FPGA while being on a modern process node makes it an appropriate FPGA on a system level for the development of energic particle instrumentation. In these systems TDCs are paired with event logic to discriminate valid events from background. With FPGA based TDCs this event logic can be embedded inside the same FPGA as the TDCs, simplifying the overall system.

Although more resolution and accuracy are always beneficial, timing requirements in modestly sized energic particle instruments are well addressed by equally modest performance TDCs; here our design target was of order 10 ps at an event rate of \textgreater 1 million events per second (MEPS) while minimizing complexity and power consumption.

\section{Implementation}
Our implementation is based on the wave union technique \cite{waveunion}, this technique measures the position of multiple edges on a tapped delay line realized in the FPGA fabric. The core idea behind this technique is to mask uncertainty of certain wide bins in the tapped delay line by having enough edges that at least one of them will be in a bin of narrow width/low uncertainty. 

For our implementation we make use of a tapped delay line composed of carry logic CCU2 blocks. We also make use of these elements for the launcher of the wave union pulse waveform. The wave union pulse waveform is simply a waveform with defined delay between the various edges that will propagate down the delay line after it is launched. A latch launches the wave union into the tapped delay line. The structure of our implementation is illustrated in Fig 1.

\begin{figure*}[!t]
  \centering
  \includegraphics[width=\textwidth]{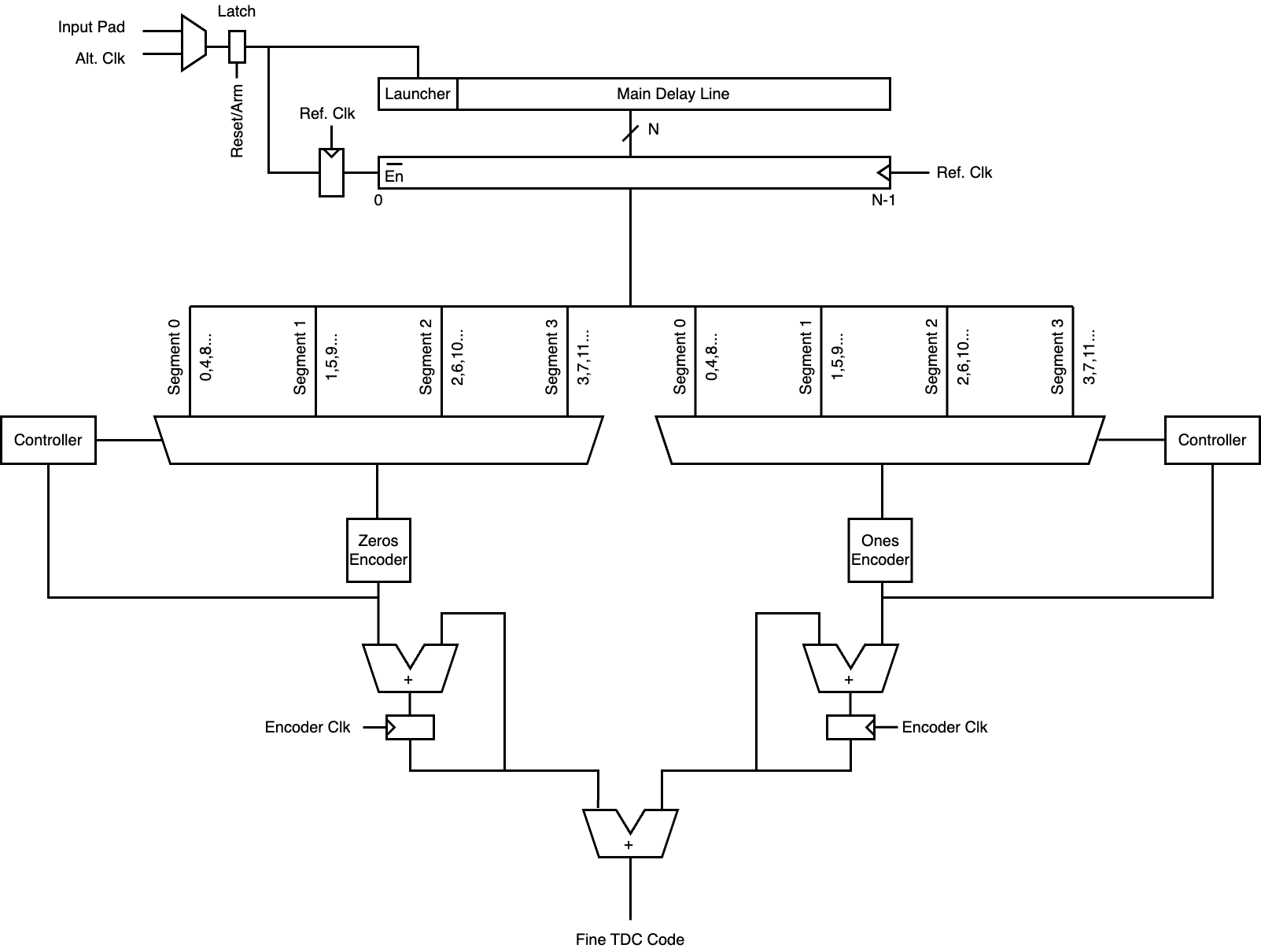}
  \caption{The TDC is started by the input latch. This starts the propagation of the wave union pulse from the launcher into the main delay line. The result is then frozen at the next reference clock edge by disabling the parallel register. The controller then adds the correct segments to calculate the total number of zeros from the right hand side of the delay line and ones from the left hand side of the delay line. The sum of these two lengths is the fine code of the TDC which interpolates within the reference clock period. The controller is started after the latch fires and provides the reset/arm signal when complete (not drawn). The coarse time is provided by a regular counter running from the reference clock.}
  \label{fig_1}
  \end{figure*}

The wave union TDC is started/launched by the input event and stopped by the clock of the system. Our reference design uses a 200 MHz clock. This clock is used by a conventional counter to provide the MSBs of the time measured with the LSBs provided by the delay line. In total the two provide an absolute time tag of the event. 

The reference clock selected to complete the TDC is of importance to the final performance. Our implementation avoided using the internal PLL in the FPGA and provided a low jitter LVDS clock running at 200 MHz. This frequency was chosen because space grade clocks are available at this frequency while it makes the tapped delay line a reasonable length in the FPGA. Higher frequencies can be used with a power/area tradeoff. The reference clock's Allen Variance should be considered over the maximum measurement window. In most ToF based particle instruments this window is very short on the order of a microsecond or less. For these short times the more common phase noise and jitter specifications can be used to select a suitable reference clock. The jitter of this clock should be less than the desired TDC resolution. For this work we used a reference clock (Skyworks: 511BBA200M000AAG) with a maximum RMS jitter of 2.1 ps.

\IEEEpubidadjcol
The inherent non-linearities in this type of FPGA based TDC require calibration. We have made both direct measurements and inferred statistical measurements of the code to time map of the TDC.

The direct measurement injects events into the TDC at known commandable times relative to the reference clock. It makes use of an Stanford Research Systems DG645 digital delay line. This instrument's clock is locked to a 10 MHz tone derived from the 200 MHz reference clock on the FPGA. Commands are issued to modify the relative delay of the event pulses at 5 ps resolution. This allows the width of each bin to be directly measured. The DG645 has a specified jitter of nominally 25 ps in the configuration used. Thus for each delay set multiple codes are returned, effectively the jitter becomes a dither. This dither is helpful to resolve the true center of each TDC code by the statistical distribution of the result. This direct measurement provides an anchor to code to time map and its associated INL and DNL, as it is effectively a independent time delay measurement. 
  
Alternatively the code to time map can be inferred by creating a histogram of events that uniformly stimulate the TDC codes \cite{WUissues}. The count of events in each bin in the histogram is then proportional to the width of the TDC bin or code. The only constraints on this method are that the INL function is independent of or weakly dependent on rate, the input pulses are uniform or white in nature, and that the integral be performed long enough that the uncertainty in counting is much less than the bin width. 

Random events meet the uniformity criteria and are naturally available in particle instruments. Alternatively, a second uncorrelated clock can be used to approximate a uniform distribution. The second clock will sweep the time of the pulse though the window uniformly with a modulus effect. No uncorrelated clock will be an exact harmonic of another, however the modulus of the two clock frequencies sets the minimum time to sweep though all of the delay line and thus the minimum event count to fully sweep through all codes. The integration time can be set by terminating the integral at the exact beat frequency of the two clocks or more trivially by integrating long enough that the histogram is statistically stationary. The later condition will be more quickly achieved when the clocks are minimally harmonic and have a large frequency modulus with respect to each other. The jitter performance of this second clock is not important and a low quality clock may be used. This clock can be routed to calibrate each TDC via a MUX before the launch latch as shown in Fig. 1.
  
With a secondary clock and the histogram based calibration method, the INL function of the TDC's tapped delay line can be extracted in a known time to a known precision on command. This process can be used when a physical variable has changed significantly enough to warrant recalibration or periodically. The time to complete this process depends on the total counts in the calibration histogram and the event rate used. For a TDC with a nominal code increment of 10 ps and 500 unique codes able to capture events at 1 MEPS, the minimum time to calibrate the system statistically to 1 ps precision would be 50 ms. This represents a very small deadtime/overhead in a realistic system. 
  
The maximum event rate possible for this type of TDC is one per clock or 200 MEPS. To achieve this event rate would require the encoder of the tapped delay line to encode the line in 1 clock cycle. To accomplish this the encoder would need to be large and clocked at 200 MHz. For our target application space this is not needed. Instead we have designed our encoder to run at 100 MHz to process events up to \textgreater 1 MEPS. This allows us to minimize the encoder area and power.
  
Near the logic transitions of the wave union in the tapped delay line glitches called bubbles can occur \cite{WUissues}. To suppress these bubbles we segment the delay line into combs. Each comb is interleaved with the other periodically (A, B, C, D, A, B, C, ...). By segmenting the tapped delay line in this manner each segment can be made free of bubbles. For our implementation we have segment the tapped delay line into 4 segments (Fig. 1). This is robust against as many as two bubbles on each transition. In experiments we have only ever observed a single bubble in the CertusPro-NX when we directly read out the tapped delay line bits. In our implementation each of these interleaved segments is further broken into linear chunks. 
  
Each chunk is processed for the number of ones or zeros from the left or right. We implement this by a priority encoder like method. The system accumulates the number of ones or zeros until an edge is detected. A multiplexer like structure feeds the two accumulators with the correct chunks based on the previous result observing that chunks will be all zeros or all ones until the last set on N chunks where N in the number of segments. The total worst case number of clock cycles to complete is N*M + K where N is the segment count, M is the chunk count per segment and K is the overhead to complete processing and store the result. In our implementation N = 4; M = 10; K = 2, and the chunk encoder is 8-bits. 

The primary delay line is composed of 137 CCU-2 elements with another 23 allocated to the launcher. This results in 320 taps as inputs to the encoder. The output code from the encoder is the simple sum of the zero and one edge distances, clocked at 100 MHz the TDC can accept 2.3 MEPS. Faster variants can be made by increasing the encoder complexity. The encoded events (fine time) and associated clock cycle (course time) are stored in a FIFO for post-processing. The base components of the TDC fit within a bounding box of 40 columns and 3 rows of PFUs, with a small number of unused slices within this box. The routing constrains make this the best measure of the effective resource utilization of the design. There are 10,850 PFUs available in the target FPGA fabric, enough for 90 TDCs of this size. In practice however further routing constrains and the need for other logic will limit the number, for this work we implemented 16 channels.

\section{Performance characterization}
\subsection{Typical Performance}
The typical performance of the TDC was measured by both the direct method and the histogram method described above. These methods produce the code to time mapping of the TDC, which can then be used to calculate the INL and DNL of the TDC. The temperature of the room was nominally 27 C.

Our implementation placed 16 TDC in the FPGA fabric. We make use of the low-power setting to generate the bitfile for the FPGA. The high performance setting results in a faster logic speed, which has better resolution but requires a longer delay line and uses more power.

Typical INL and DNL functions are shown in Fig 2. The typical code width is 10.9 ps, this is the average resolution or LSB. The INL has a standard deviation of nominally 1 LSB and a max error +/- 3 LSB. The DNL is bounded by (+1.13,-0.77) LSB, thus it is monotonic but has positive errors slightly larger than the resolution at some codes.

\begin{figure}[!t]
  \centering
  \includegraphics[width=3.5in]{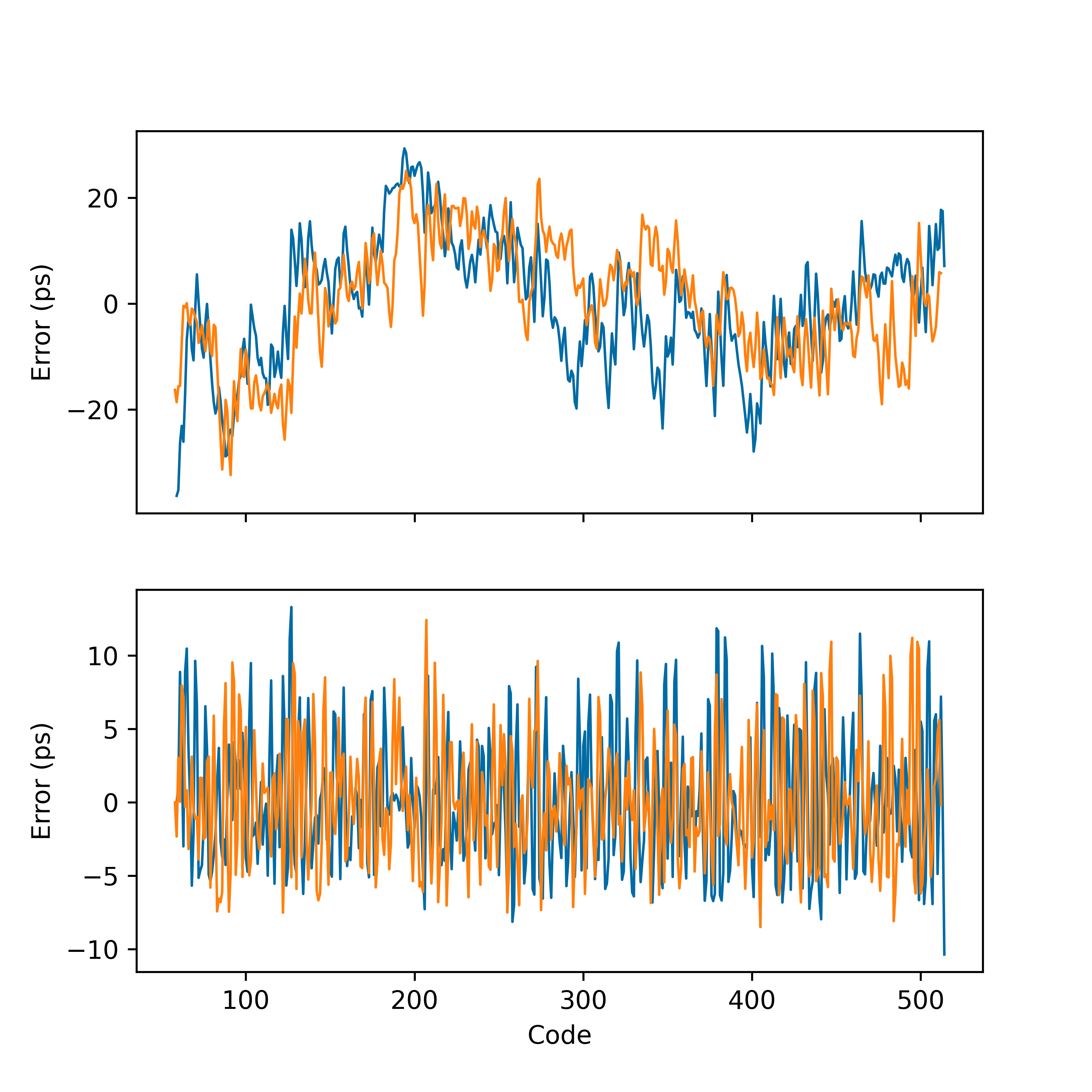}
  \caption{The INL (top) and DNL (bottom) functions of two representative TDCs}
  \label{fig_2}
  \end{figure}

\subsection{Power Consumption}
The FPGA is provided with power rails at 3.3, 1.8 and 1.0 V. The 3.3 V rail also powers 2 clocks and 2 memories, these typically consume 33 mA. With a single TDC implemented the system consumes 42, 8, and 78 mA on the 3.3, 1.8 and 1.0 V supplies respectively. Each additional TDC consumes approximately 6 mW of power more on the 1.0 V rail. With 16 TDCs the system consumes 42, 8, 169 on the 3.3, 1.8 and 1.0 V supplies respectively. With the implementation used there are no significant change in power consumption as a function of event rate. This is due to the relatively low activity rate of events (1 MEPS max) in comparison with the reference clock (200 MHz). Total power consumption is approximately 225 mW + 6 mW/channel. For systems that combined TDCs with event logic the incremental power consumption of 6 mW/channel is the most appropriate metric for estimating power consumption for adding TDCs; for stand alone TDC use cases the total power is the more appropriate metric.

\subsection{Voltage Dependence}
The TDC fabric and buffers propagation delays are dependent on the supply voltages. The core voltage (1.0 V) is the primary voltage that determines the delay line speed, while the auxiliary voltage (1.8 V) and I/O voltages (3.3 V) impact buffers and other logic. 

We measured TDC code to time map while varying these voltages. From this we extracted the slopes and offsets of the maps to determine the voltage sensitivity. The TDC was most sensitive to core voltage (1.0 V) changes with a fractional change of nominally 0.22\%/mV. This a 1 mV change in the core voltage will result in a 1 LSB change at the highest codes of the TDC. Sensitivity of the auxiliary voltage (1.8 V) is lower at 0.027\%/mV (8 mV for 1 LSB). There was minimal sensitivity on the I/O voltage (3.3 V), however the input buffer voltage creates a shift. This is nominally 11 mV per LSB in delay. 

The voltage sensitivity implies great care is required with the power supplies to the FPGA to achieve the best possible performance. The core voltage should not vary over the time window of interest by more than 1 mV to achieve maximal accuracy. These effects are most important for absolute time measurements using a single channel. For differential measurements the effective window of interest becomes a short time between channels, which greatly reduces the effect. 

If it is not practical to construct a system with low voltage ripple on the FPGA supplies an alternative system level approach not implemented here is to create a ratiometric TDC. This can be implemented by measuring the same event twice on a longer or looped delay line with 1 clock cycle between the measurements. The code difference between the two measurements represents the velocity on the tapped line while either of the measurements represents the interpolated time. Variance in the core voltage effectively changes the velocity of the line so this method would remove it. We have not elected to do that here to save power and area. 

\subsection{Temperature Dependence}
The performance of the TDC as a function of temperature is dependents on three factors: (1) the changes in propagation time of the FPGA fabric due to temperature, (2) the change in the reference clock frequency as a function of temperature, (3) changes in the supply voltages as a function of temperature. The reference clock used has a specified temperature stability of +/- 25 ppm over -40 to 85 C. This change is negligible compared to the LSB of the TDC and can be ignored. The supply voltages will change with the reference voltage temperature coefficient in the regulator and the passives that set the voltage. We used LT3080 regulators to provide the supply voltages with a 10 ppm/C resistor used to provide the set point. Given the voltage sensitivity of the TDC a 1 mV change on the 1.0 V core supply should be measurable (0.1\%). From the resistor alone this would occur at 100 C change in temperature. Meanwhile the current in the LT3080 set pin changes nominally 0.5\% in 50 C. For the core voltage this is 1 mV in 10 C, which implies that voltage shifts from temperature changes could become important in nominal terms at 10 C temperature shifts, largely due to the regulator reference. The changes to the FPGA fabric itself needs to be experimentally determined. 

To test the overall system we placed the FPGA test card a thermal chamber and measured the INL as a function of temperature. This estimates the residual error that would occur if the system is not recalibrated. Fig. 4 shows the residuals of the observed INL function at 3 temperatures. 34.1 C was used as the reference temperature. The INL functions were extracted by the histogram method, which would typically be used for self-calibration. The standard deviation of the residual of the INL function is only 0.41 ps at the reference temperature but increases 2.5 ps at a nominal temperature offset of 5 C. The temperature of the FPGA was measured by the internal temperature measurement provided by the ADC unit in the FPGA. Based on these results and the LSB of 10.9 ps for this design it is conservative to recalibrate the system when the temperature has changed by more than 5-10 C. This temperature stability is inline with expectations based on the component contribution. 

\begin{figure}[!t]
  \centering
  \includegraphics[width=3.5in]{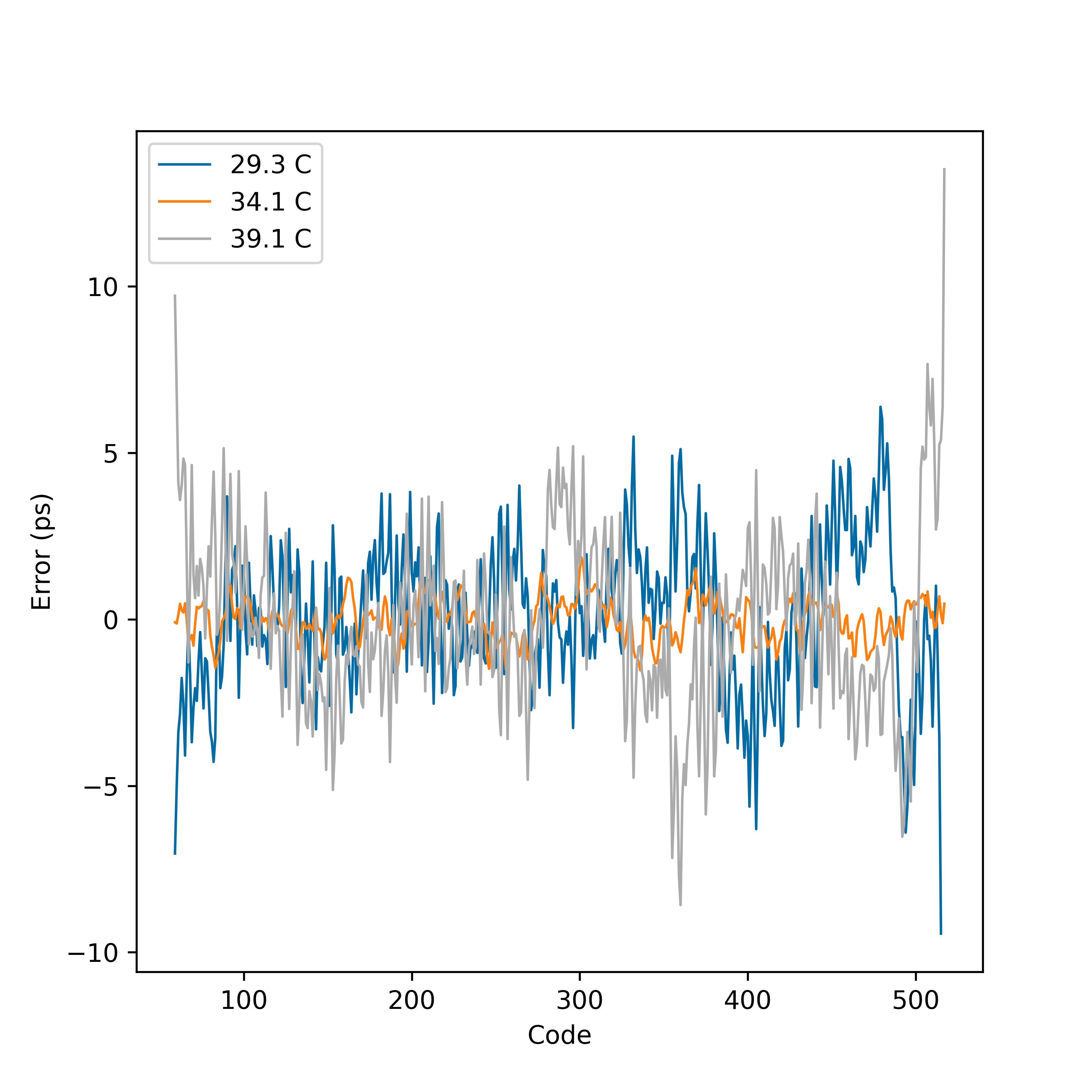}
  \caption{Residuals between INL functions extracted by the histogram method with a reference INL function taken at 34.1 C.}
  \label{fig_3}
  \end{figure}

\subsection{Differential time measurements}
For ToF system it is the time difference between multiple events that is of interest. In this case the absolute times recorded by multiple TDCs will be differenced appropriately by the event logic to determine if an event of interest has occurred.

To simulate this use case we ran the DG645 delay generator unlocked from the FPGA clock with a commanded delay between the two output edges input into two TDC channels, while varying the time between the two edges. We varied the time between the two edges over a range of 1000 ns in steps of 25 ns. We then recorded 512 trials for each delay. Examples of this data are shown for 3 delay times in Fig. 4.

\begin{figure}[!t]
  \centering
  \includegraphics[width=3.5in]{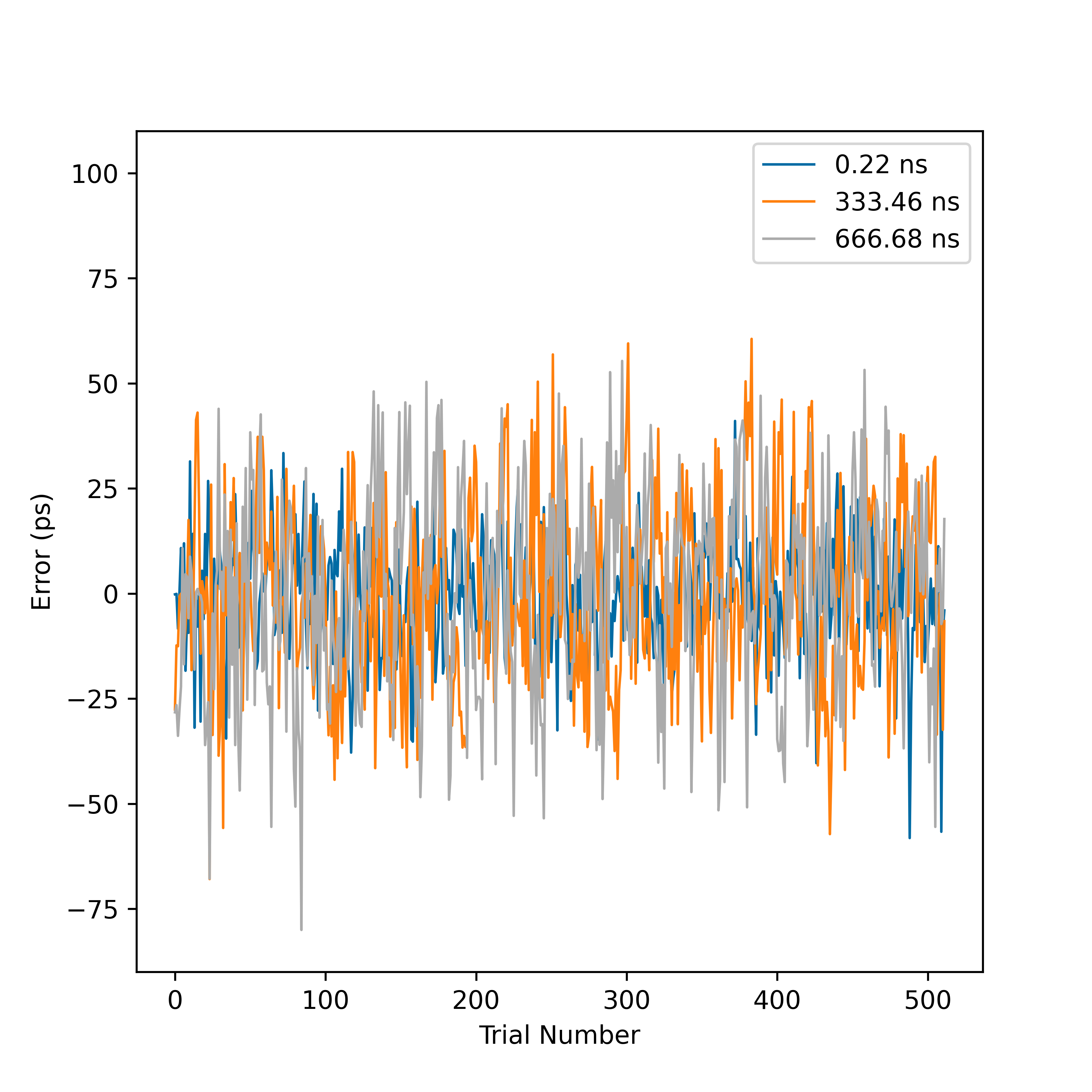}
  \caption{Differential time measurement between two TDC channels at 0.22 ns, 333.46 ns and 666.68 ns}
  \label{fig_4}
  \end{figure}

The standard deviation of these trials contains the jitter of the DG645, the reference clock, and the two TDC's random errors. We observed an error bounded by 22 ps over the 1000 ns window. The jitter between outputs on the DG645 is specified to be less than 15 ps for short times and is typically \textless 12 ps. To assess the inherent jitter in a single TDC channel we looped back the reference clock with a cable into a TDC channel. This resulted in a standard deviation of 0.33 LSB or 3.6 ps. Meanwhile the reference clock has a specified jitter of 2.1 ps. The quadrature of these errors is nominally 16 ps. This suggests that the errors might not be completely uncorrelated or some noise is getting into the measurement in the differential configuration that is not fully suppressed. This can be the ripple of the power supply in the time between the first and second edge measurements for instance. We feel is is conservative to remove 10 ps of error from the observed figure to represent the DG645 contribution to place an upper bound on the typical random error of a differential measurement. This is a typical 1-sigma error \textless 20 ps. The absolute upper bound is the observed 22 ps and the best reasonable case is 16 ps. 

\section{Conclusion}

We have designed a low power TDC in the FPGA fabric of a low power radiation tolerant FPGA. The performance of the TDC is sufficient to support a variety of ToF and timing based instrumentation for space applications. The TDC can in principle be integrated with event or other instrument logic to provide a compact and resource efficient implementation. A full instrument demonstration with such integration is the next step to identify any systematic effects beyond the initial characterization performed here. Based on the work here we believe it is realistic to achieve 1-sigma absolute timing errors of nominally 20 ps.

\vspace{11pt}




\vfill

\end{document}